\documentclass[aps,prb,preprint,superscriptaddress,showpacs]{revtex4}

\usepackage{dcolumn}
\usepackage{bm}
\usepackage{amsmath} 
\usepackage{amsfonts} 
\usepackage[english]{babel}  
\usepackage[T1]{fontenc}
\usepackage{graphicx}
\usepackage{epstopdf}

\begin{document}

\preprint{Phys.\ Rev.\ B Rapid Communications (In Press)}

\title{Charge-transfer excitons in strongly coupled organic semiconductors}

\author{Jean-Fran\c{c}ois Glowe}

\author{Mathieu Perrin}
\affiliation{D\'epartement de physique et Regroupement qu\'eb\'ecois sur les mat\'eriaux de pointe, Universit\'e de Montr\'eal, C.P.\ 6128, Succursale centre-ville, Montr\'eal (Qu\'ebec) H3C~3J7, Canada}

\author{David Beljonne}
\affiliation{Service de Chimie des Mat\'eriaux Nouveaux, Universit\'e de Mons-Hainaut, Place du Parc 20, B-7000 Mons, Belgium}

\author{Sophia C.\ Hayes}
\affiliation{Department of Chemistry, University of Cyprus, P.O.\ Box 20537, 1678 Nicosia, Cyprus}

\author{Fabrice Gardebien}
\affiliation{Service de Chimie des Mat\'eriaux Nouveaux, Universit\'e de Mons-Hainaut, Place du Parc 20, B-7000 Mons, Belgium}

\author{Carlos Silva}
\email[Electronic address: ]{carlos.silva@umontreal.ca}
\affiliation{D\'epartement de physique et Regroupement qu\'eb\'ecois sur les mat\'eriaux de pointe, Universit\'e de Montr\'eal, C.P.\ 6128, Succursale centre-ville, Montr\'eal (Qu\'ebec) H3C~3J7, Canada}

\date{\today}

\begin{abstract}
Time-resolved and temperature-dependent photoluminescence measurements on one-dimensional sexithiophene lattices reveal intrinsic branching of photoexcitations to two distinct species: self-trapped excitons and dark charge-transfer excitons (CTX; $\gtrsim 5$\% yield), with radii spanning 2--3 sites. The significant CTX yield results from the strong charge-transfer character of the Frenkel exciton band due to the large free exciton bandwidth ($\sim400$\,meV) in these supramolecular nanostructures.
\end{abstract}


\maketitle

The physics of organic semiconductor materials attracts enormous multidisciplinary interest due to emerging applications in optoelectronics. Their electronic properties depend sensitively upon supramolecular structure. A promising strategy to enhance carrier mobilities, for example, is to induce supramolecular order in configurationally disordered systems such as solution-processable conjugated polymers.\cite{Sirringhaus2005} Prototypical examples of this approach are regioregular polythiophenes, which display field-effect mobilities up to 0.6\,cm$^2$\,V$^{-1}$\,s$^{-1}$ in film microstructures showing lamellar interchain packing,~\cite{Sirringhaus1999,Mcculloch2006,DeLongchamp2007} resulting in two-dimensional delocalization of charge carriers.~\cite{Osterbacka2000,Chang2006} Co-facial interactions between polymer chains lead to H-aggregates with spatially correlated energetic disorder.~\cite{Spano2005} In the film microstructures that result in the best field-effect mobilities the supramolecular electronic (resonant Coulomb) coupling energies $J$ are $\sim 30$\,meV,~\cite{Clark:2007, Clark:2009} which are weak compared to molecular reorganization energies ($\sim 180$\,meV). This is a result of long conjugation lengths in the lamellar architecture, such that local site interactions are weak.~\cite{Meskers2000} As the chain length decreases below the conjugation length, $J$ increases and may enter an `intermediate' regime.~\cite{Manas1998,Beljonne2000,Westenhoff:2006,Barford2007,Gierschner2008} Here, we address the nature of primary photoexcitations in this regime. We demonstrate, by means of time and temperature-dependent photoluminescence (PL) measurements on chiral, helical sexithiophene stacks~\cite{Wolffs08} (labelled T6 for brevity), that excitation of the H-aggregate band with femtosecond laser pulses produces a high intrinsic yield ($\gtrsim 5$\%) of charge-transfer excitons (CTX), in which the center of mass of electron and hole are localized at different sites of the stack. These are dark states, which recombine to populate luminescent states with a distribution of rate constants, and we determine that their radius is confined to 2--3 sites. The direct CTX yield is a consequence of the large exciton bandwidth, providing access to charge-transfer states.

T6 (99.9\% purity~\cite{Wolffs08}) solutions of  $10^{-4}$\,M in anhydrous n-butanol were studied in a temperature-controlled UV-grade fused silica cuvette (1-mm pathlength). Films were produced by drop-casting the solution on Spectrosil substrates. Absorption spectra were measured with a Varian Cary-500 spectrometer. Time-resolved PL measurements were performed with a femtosecond laser system (KMLabs Dragon, 780\,nm, 40\,fs FWHM, 1\,kHz repetition rate, 1.4\,mJ/pulse), which was frequency-doubled in a $\beta$-BBO crystal to generate 390-nm (3.19-eV) pulses, and a spectrograph with a gated, intensified CCD camera (Princeton Instruments SP-2156 and PIMAX 1024HB). Femtosecond absorption transients were measured with this ultrafast source, and probing with a white-light continuum generated in a CaF$_2$ window. 


\begin{figure}
\includegraphics[width=7.5cm]{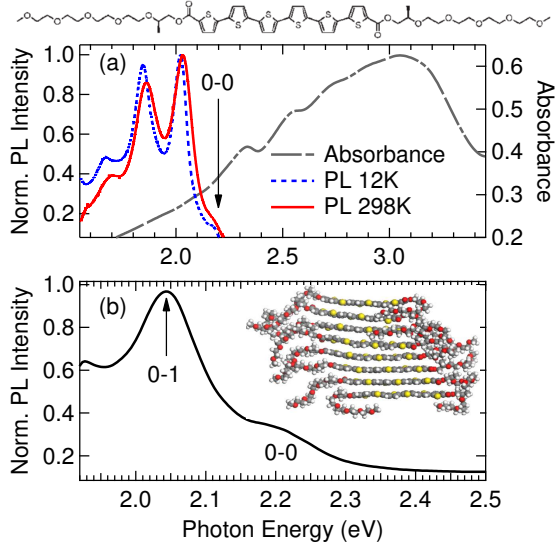}
\caption{(color online) Molecular structure of T6. (a) Absorbance spectrum of T6 nanostructures  in $n$-butanol solution at 283\,K, and steady-state PL spectra of a drop-cast film at 12 and 298\,K. (b) Delayed PL spectrum from 6.5\,$\mu$s after excitation and a gate width of 500\,$\mu$s  in $n$-butanol solution. The inset depicts the chiral, co-facial T6 stacks.}
\label{fig1}
\end{figure}

Spontaneous supramolecular organization of T6 (Fig.~\ref{fig1}) into chiral, columnar stacks is observed with thermotropic reversibility~\cite{Henze2006} below a transition temperature of ($313 \pm 12$)\,K at the solution concentration used in this study. The supramolecular packing and the size distribution of these nanostructures, which grow by a nucleation self-assembly mechanism, depends sensitively upon the material purity and on the solution cooling protocol.~\cite{Wolffs08} In all of the studies reported here, the sample was cooled at a rate of $\sim 1$\,K\,min$^{-1}$, but the PL spectral band shape and the time-resolved PL dynamics do not depend sensitively upon the cooling protocol. Fig.~\ref{fig1}(a) displays the solution absorption spectrum in the supramolecular phase (283\,K). We have previously extracted a free-exciton bandwidth $W = 4J \approx 400$\,meV by analysing this spectrum,~\cite{Westenhoff:2006} which places these nanostructures firmly in the `intermediate' electronic coupling regime.~\cite{Meskers2000} Also shown are the corresponding steady-state PL spectra of a drop-cast film at room and low temperature. These display a similar spectral band shape as in cold solution,~\cite{Westenhoff:2006} indicating that we preserve the supramolecular structure upon casting the film. The origin (0--0) vibronic band is weak due to the H-aggregate nature of the architecture,~\cite{Spano2007a} and is only visible due to energetic disorder.~\cite{Spano2009} Its weak temperature dependence is consistent with the large $W$ deduced from the absorption spectrum; even at room temperature, thermal excitation is not sufficient to partially allow this vibronic feature. This is unlike the case in polythiophene films, which feature much smaller $W$, and are in a weak excitonic coupling regime.~\cite{Clark:2007, Clark:2009} Quantum chemical calculations of related supramolecular nanostructures indicate that the PL spectrum is due to vibrationally dressed Frenkel excitons with a highly localized center-of-mass due to a disorder width of several hundred meV and low spatial correlation.~\cite{Spano2007a,Spano2009,Spano2008} Fig.~\ref{fig1}(b) shows the delayed PL spectrum over $\mu$s timescales,  which is identical to the steady-state spectra in part (a), and displays linear integrated intensity below excitation fluences of $\sim~800$\,$\mu$J\,cm$^{-2}$.~\cite{SuppInfo}

\begin{figure}
\includegraphics[width=7.5cm]{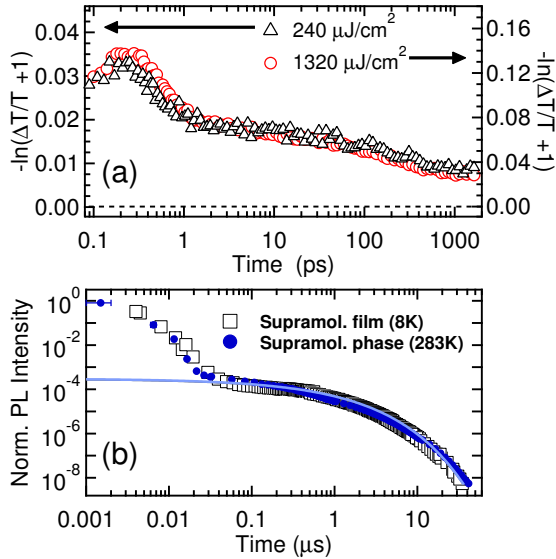}
\caption{(color online) (a) Femtosecond absorption transients probing at 1.42\,eV with two pump fluences indicated in the caption. (b) Time-resolved PL intensity for T6 in the supramolecular phase (blue filled circles) in solution, and in a drop-cast film of the stacks at 8\,K (black open squares). The curves through the delayed PL decay are stretched-exponential fits with $\beta = 0.5$ and $\tau = 300 \pm 50$\,ns for the solution at 283\,K and for the film.}
\label{fig3}
\end{figure}

We now consider photoexcitation dynamics in T6 lattices by comparing the $1B_u$ femtosecond transient absorption (Fig.~\ref{fig3}(a)) and time-dependent PL measurements over microsecond time windows (Fig.~\ref{fig3}(b)).  All decay behavior is independent of fluence (as demonstrated in part (a)). Following a rapid initial relaxation ($< 1$\,ps) the exciton decay is biphasic over a microsecond time window. We observe an exponential decay with $\sim 1$\,ns time constant, characteristic of exciton lifetimes in conjugated oligomer nanostructures.~\cite{Daniel:2005, Westenhoff:2006} We observe a concomitant delayed PL signal (blue circles) which is spectrally identical to the steady-state PL (Fig.~\ref{fig1}); the delayed PL is therefore excitonic. The spectrally integrated, delayed PL intensity is linear over the pump fluence range investigated,~\cite{SuppInfo} and the decay kinetics are independent of temperature (see below), ruling out bimolecular recombination events such as triplet-triplet annihilation as the origin of this delayed PL. We therefore attribute it to recombination of CTXs, analogous to geminate polaron pairs in conjugated polymer films.~\cite{Arkhipov2004} 

The functional form of the delayed PL decay is stretched exponential,
$(I_{0} \exp [- ( t / \tau) ^{\beta} ] )$, 
with $ \beta = 0.5 $ and $\tau = 300 \pm 50$\,ns. Fig.~\ref{fig3}(b) displays this function. 

 In order to explore the origin of the non-exponential decay of the delayed PL, we present related measurements in a drop-cast film of supramolecular stacks at 8\,K (open squares in Fig.~\ref{fig3}(b); note that these data are offset for clarity), in which the PL spectra are consistent with those in Fig.~\ref{fig1}. The delayed PL decay follows a stretched exponential with similar parameters as the supramolecular phase in solution, ruling out a time-dependent rate constant  due to endothermic activation. We therefore ascribe the nonexponential PL decay simply to a distribution of rate constants $ \gamma$.

The relative contribution of the delayed PL to the time-integrated PL intensity provides a lower limit to the CTX photogeneration yield, $ \eta $. We extract $\eta = 4.8 \pm 1.0$\% in the supramolecular phase in solution. In contrast, we measure $ \eta = 0.25 \pm 0.15$\% at 350\,K, well above the transition temperature for disassembly. In the stack, $\eta$ does not depend strongly upon temperature down to 8\,K, where we find $3.9 \pm 1.0$\%, indicating that CTXs are produced directly by photoexcitation. We propose the following photophysical picture. Initial photoexcitation of the Frenkel free-exciton band branches into highly localised (self-trapped) excitons,~\cite{Spano2007a} displaying prompt exponential dynamics, in which the electron and hole centers of mass are delocalised over essentially one oligomer lattice site. Concomitantly, CTXs are produced directly with $\gtrsim 5$\% efficiency. The branching occurs due to resonance between Frenkel excitons and charge-transfer (CT) states, which normally lie 0.2--0.3\,eV above localized Frenkel exciton states,~\cite{Psiachos2009} because the exciton bandwidth in these strongly coupled stacks is large enough to encompass the CT states.~\cite{Scholes2008} CTXs then recombine with a distribution of rate constants resulting in the non-exponential delayed PL decay.

\begin{figure}
\includegraphics[width=7.5cm]{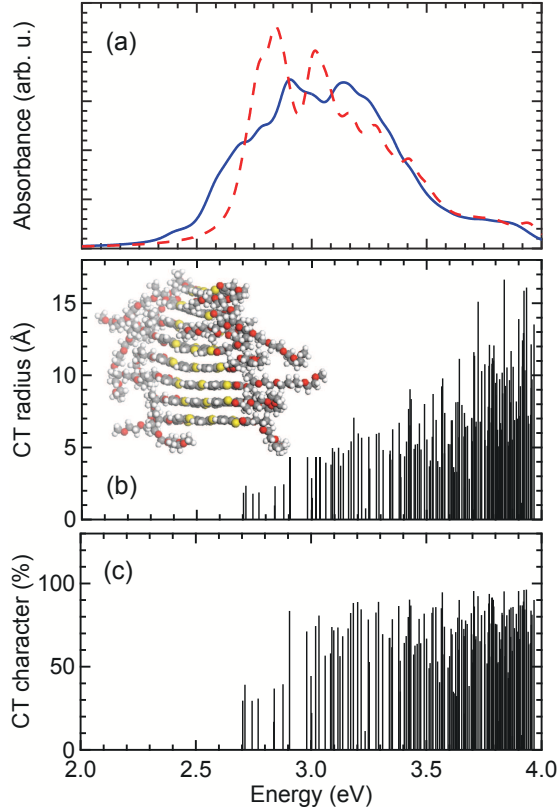} 
\caption{(color online) (a) Ensemble (solid line) and single-stack (dashed line) absorption spectra of a stack of 8 T6 simulated at the INDO/SCI level.~\cite{Henze2006} Lorentzian functions with linewidth 0.05\,eV have been used to convolute the spectra. The ensemble spectra are obtained by averaging over 20 snapshots extracted along the MD trajectory. (b)  Charge-transfer radius versus vertical excitation energy for one snapshot (see inset), and (c) its fractional charge-transfer character.}
\label{david}
\end{figure}

We have performed quantum chemical calculations on a stack of 8 T6 molecules. These extend previous force-field molecular dynamics (MD) simulations.~\cite{Henze2006}  Fig.~\ref{david}(a) displays the ensemble absorption spectrum obtained by averaging 20 snapshots extracted along the MD trajectory (solid line), and that in a single-snapshot (dotted line), simulated at the INDO/SCI level.~\cite{IndoNote} We calculate the CT radius (for which the spatial extent of the wavefunction spans more than one oligomer, see ref.~\onlinecite{Scholes2008}) as a function of the vertical excitation energy (Fig.~\ref{david}(b) for one stack snapshot). We also plot the fractional CT character versus excitation energy in Fig.~\ref{david}(c). We note that the CT character and the corresponding radius increase with energy, and that it dominates above $\sim 3$\,eV.

\begin{figure}
\includegraphics[width=7.5cm]{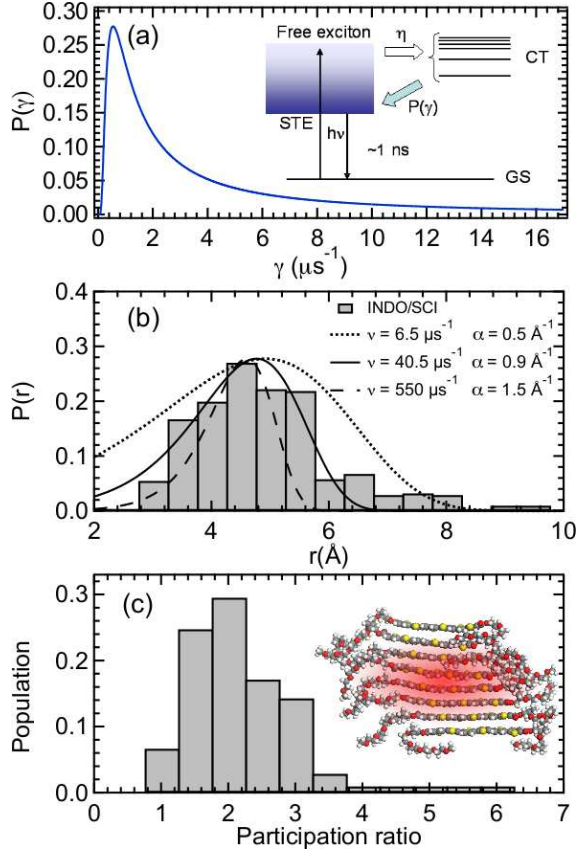}
\caption{(color online) (a) Distribution of recombination rate constants given by equation~\ref{Eq:Pgamma}. The inset depicts the photophysical model put forth here. Vertical excitation of the free exciton band is followed by branching to self-trapped (Frenkel) excitons (STE, 95\%) and charge-transfer (CTX, 5\%). The latter recombine to the former state with $P(\gamma)$. 
(b) INDO/SCI histogram of charge-transfer radii built on the basis of an ensemble of 20 structures extracted from MD simulations. All excited states with dominant ($>50$\%) CT character that are located below 3.3\,eV have been included. The curves show fits to the exponential distance dependence of the rate constant. 
(c) Participation ratio (measure of excited-state delocalization) for the ensemble of 20 snapshots (b). The inset depicts the spatial extent of exciton.}
\label{fig4}
\end{figure}

A PL decay function characterized by a distribution of decay constants $ P(\gamma) $ is its Laplace transform,
\begin{equation}
I(t) = I_0 \int_{0}^{\infty} P(\gamma) \exp \left( - \gamma t \right)\,\mathrm{d}\gamma
\label{Eq:Laplace}
\end{equation}
where $\int_{0}^{\infty} P(\gamma)\,\mathrm{d}\gamma = 1$. For a stretched exponential decay with $ \beta = 0.5 $, the probability distribution is
\begin{equation}
P(\gamma) = \frac{\exp \left[ - \left( 4 \gamma \tau \right)^{-1} \right] }{\sqrt{4 \pi \gamma^{3} \tau}}
\label{Eq:Pgamma}
\end{equation}
and is shown in Fig.~\ref{fig4}(a) with $ \tau = 300$\,ns. This distribution is centered at $ \gamma_{\mathrm{peak}} = (0.55 \pm 0.1)$\,$\mu$s$^{-1}$.

We postulate that $P(\gamma)$ arises from a distribution of CTX radii $P(r)$. Fig.~\ref{fig4}(b) displays a histogram of $P(r)$ based on the quantum-chemical calculations described above, and Fig.~\ref{fig4}(c) plots the exciton participation ratio. This quantity displays the distribution of the population in which the electron and hole center of mass are separated by a specific number of lattice sites, determined from the calculations presented in Fig.~\ref{david}, and expresses the extent of exciton delocalization. $P(\gamma)$ is temperature independent, therefore recombination must occur via a tunneling mechanism. The electron-hole electronic coupling matrix element decreases exponentially with distance due to the exponential radial character of the electronic wave functions. If $\gamma$ follows a golden rule, then $ \gamma(r) = \nu \exp \left( -\alpha r \right) $.  We hence plot $P(r)$ in \ref{fig4}(b) for various values of $\gamma_{0}$ and $\alpha$, and find the best agreement with the quantum-chemical calculations for $ \gamma_{0} = 40.5$\,$\mu$s$^{-1}$ and $\alpha = 0.9$\,\AA$ ^{-1}$. We note that $\gamma_{\mathrm{peak}}$ is comparable to rates measured in donor-bridge-acceptor triads~\cite{Barbara1996} but is surprisingly low for a co-facial $\pi$-stack. We speculate that the polar environment surrounding the $\pi$ stack defined by the oligoethyleneoxide end-groups plays a role by stabilizing the CTX. Such enivronment may also play a role in establishing $\eta$.

The high value of $ \eta $ is in contrast to that in configurationally disordered conjugated polymer films, where $\eta \ll 1$\%.~\cite{Silva2002} The magnitude of $J$ is important in this respect. In such co-facial aggregates with strong excitonic coupling, intermolecular interactions enable quasi-particle energy dispersion supporting CT states. In T6 stacks a large free exciton bandwidth $W$ renders an intermolecular CT state accessible. As $W$ decreases with increasing $\pi$-conjugation length,~\cite{Westenhoff:2006} the influence of CT states decreases along with $\eta$.  Nevertheless, weakly-allowed CTX states are important in a complete theoretical description of photoinduced absorption spectra of conjugated polymers in the solid state.~\cite{Wang2008,Psiachos2009}


We have demonstrated that in organic semiconductors, the supramolecular coupling energy dominates the nature of the primary photoexcitations. The large free exciton bandwidth is significantly larger than attainable in the most highly organized semiconductor polymer microstructures,~\cite{Clark:2007} but the primary photoexcitations are highly localized.  The Frenkel exciton band mixes with charge-transfer states, which play an important role in the primary photophysics. This work is consistent with independent experimental evidence for discrete, localized CTX states at the surface of  pentacene crystalline films, probed by means of time-resolved two-photon photoemission spectroscopy by exciting well in the LUMO band.~\cite{Muntwiler2008} We consider that these conclusions are of general importance for a detailed description of the electronic structure of organic semiconductors.

\begin{acknowledgments}
We acknowledge gratefully the  collaboration involving the groups of Albert Schenning and Jim Feast for the the synthesis of T6, and particularly Martin Wolffs, for the purification of this material. CS is funded by NSERC, CFI, and the CRC Program. The work in Mons is financially supported by the Belgian National Science Foundation (FNRS FRFC No.\ 2.4560.00), by the EC STREP project MODECOM (NMP-CT-2006-016434), and by the Belgian Federal Science Policy Office in the framework of the ``P\^{o}le d'Attraction Interuniversitaire en Chimie Supramol\'{e}culaire et Catalyse Supramol\'{e}culaire (PAI 5/3)''. DB is a Research Director of the FNRS.
\end{acknowledgments}


\begin{thebibliography}{31}
\expandafter\ifx\csname natexlab\endcsname\relax\def\natexlab#1{#1}\fi
\expandafter\ifx\csname bibnamefont\endcsname\relax
  \def\bibnamefont#1{#1}\fi
\expandafter\ifx\csname bibfnamefont\endcsname\relax
  \def\bibfnamefont#1{#1}\fi
\expandafter\ifx\csname citenamefont\endcsname\relax
  \def\citenamefont#1{#1}\fi
\expandafter\ifx\csname url\endcsname\relax
  \def\url#1{\texttt{#1}}\fi
\expandafter\ifx\csname urlprefix\endcsname\relax\def\urlprefix{URL }\fi
\providecommand{\bibinfo}[2]{#2}
\providecommand{\eprint}[2][]{\url{#2}}

\bibitem[{\citenamefont{Sirringhaus}(2005)}]{Sirringhaus2005}
\bibinfo{author}{\bibfnamefont{H.}~\bibnamefont{Sirringhaus}},
  \bibinfo{journal}{Adv. Mater.} \textbf{\bibinfo{volume}{17}},
  \bibinfo{pages}{2411} (\bibinfo{year}{2005}).

\bibitem[{\citenamefont{Sirringhaus et~al.}(1999)\citenamefont{Sirringhaus,
  Brown, Friend, Nielsen, Bechgaard, Langeveld-Voss, Spiering, Janssen, Meijer,
  Herwig et~al.}}]{Sirringhaus1999}
\bibinfo{author}{\bibfnamefont{H.}~\bibnamefont{Sirringhaus}}
  \bibnamefont{et~al.}, \bibinfo{journal}{Nature}
  \textbf{\bibinfo{volume}{401}}, \bibinfo{pages}{685} (\bibinfo{year}{1999}).

\bibitem[{\citenamefont{Mcculloch et~al.}(2006)\citenamefont{Mcculloch, Heeney,
  Bailey, Genevicius, Macdonald, Shkunov, Sparrowe, Tierney, Wagner, Zhang
  et~al.}}]{Mcculloch2006}
\bibinfo{author}{\bibfnamefont{I.}~\bibnamefont{McCulloch}}
  \bibnamefont{et~al.}, \bibinfo{journal}{Nature Mater.}
  \textbf{\bibinfo{volume}{5}}, \bibinfo{pages}{328} (\bibinfo{year}{2006}).


\bibitem[{\citenamefont{DeLongchamp et~al.}(2007)\citenamefont{DeLongchamp,
  Kline, Lin, Fischer, Richter, Lucas, Heeney, McCulloch, and
  Northrup}}]{DeLongchamp2007}
\bibinfo{author}{\bibfnamefont{D.~M.} \bibnamefont{DeLongchamp}}
 \bibnamefont{et~al.}, \bibinfo{journal}{Adv. Mater.}
  \textbf{\bibinfo{volume}{19}}, \bibinfo{pages}{833} (\bibinfo{year}{2007}).

\bibitem[{\citenamefont{\"{O}sterbacka
  et~al.}(2000)\citenamefont{\"{O}sterbacka, An, Jiang, and
  Vardeny}}]{Osterbacka2000}
\bibinfo{author}{\bibfnamefont{R.}~\bibnamefont{\"{O}sterbacka}}
 \bibnamefont{et~al.},
  \bibinfo{journal}{Science} \textbf{\bibinfo{volume}{287}},
  \bibinfo{pages}{839} (\bibinfo{year}{2000}).

\bibitem[{\citenamefont{Chang et~al.}(2006)\citenamefont{Chang, Clark, Zhao,
  Sirringhaus, Breiby, Andreasen, Nielsen, Giles, Heeney, and
  McCulloch}}]{Chang2006}
\bibinfo{author}{\bibfnamefont{J.-F.} \bibnamefont{Chang}}
 \bibnamefont{et~al.},
  \bibinfo{journal}{Phys. Rev. B} \textbf{\bibinfo{volume}{74}},
  \bibinfo{pages}{115318} (\bibinfo{year}{2006}).

\bibitem[{\citenamefont{Spano}({2005})}]{Spano2005}
\bibinfo{author}{\bibfnamefont{F.~C.}~\bibnamefont{Spano}}, \bibinfo{journal}{J.
  Chem. Phys.} \textbf{\bibinfo{volume}{122}},  \bibinfo{eid}{115318} (\bibinfo{year}{2005}); 
\textbf{\bibinfo{volume}{126}}, \bibinfo{eid}{159901} (\bibinfo{year}{2007}).

\bibitem[{\citenamefont{Clark et~al.}(2007)\citenamefont{Clark, Silva, Friend,
  and Spano}}]{Clark:2007}
\bibinfo{author}{\bibfnamefont{J.}~\bibnamefont{Clark}}
 \bibnamefont{et~al.},
  \bibinfo{journal}{Phys. Rev. Lett.} \textbf{\bibinfo{volume}{98}},
  \bibinfo{eid}{206406} (\bibinfo{year}{2007}).

\bibitem[{\citenamefont{Clark et~al.}(2009)\citenamefont{Clark,, Friend,
 Spano, and Silva}}]{Clark:2009}
\bibinfo{author}{\bibfnamefont{J.}~\bibnamefont{Clark}}
 \bibnamefont{et~al.},
  \bibinfo{journal}{Appl. Phys. Lett.} \textbf{\bibinfo{volume}{93}},
  \bibinfo{eid}{163306} (\bibinfo{year}{2009}).
  
\bibitem[{\citenamefont{Meskers et~al.}(2000)\citenamefont{Meskers, Janssen,
  Haverkort, and Wolter}}]{Meskers2000}
\bibinfo{author}{\bibfnamefont{S.~C.~J.} \bibnamefont{Meskers}}
 \bibnamefont{et~al.}, \bibinfo{journal}{Chem. Phys.}
  \textbf{\bibinfo{volume}{260}}, \bibinfo{pages}{415} (\bibinfo{year}{2000}).

\bibitem[{\citenamefont{Manas and Spano}(1998)}]{Manas1998}
\bibinfo{author}{\bibfnamefont{E.~S.} \bibnamefont{Manas}} \bibnamefont{and}
  \bibinfo{author}{\bibfnamefont{F.~C.} \bibnamefont{Spano}},
  \bibinfo{journal}{J. Chem. Phys.} \textbf{\bibinfo{volume}{109}},
  \bibinfo{pages}{8087} (\bibinfo{year}{1998}).

\bibitem[{\citenamefont{Beljonne et~al.}(2000)\citenamefont{Beljonne, Cornil,
  Silbey, Millie, and Bredas}}]{Beljonne2000}
\bibinfo{author}{\bibfnamefont{D.}~\bibnamefont{Beljonne}}
 \bibnamefont{et~al.},
  \bibinfo{journal}{J. Chem. Phys.} \textbf{\bibinfo{volume}{112}},
  \bibinfo{pages}{4749} (\bibinfo{year}{2000}).

\bibitem[{\citenamefont{Westenhoff et~al.}(2006)\citenamefont{Westenhoff,
  Abrusci, Feast, Henze, Kilbinger, Schenning, and Silva}}]{Westenhoff:2006}
\bibinfo{author}{\bibfnamefont{S.}~\bibnamefont{Westenhoff}}
 \bibnamefont{et~al.},
  \bibinfo{journal}{Adv. Mater.} \textbf{\bibinfo{volume}{18}},
  \bibinfo{pages}{1281} (\bibinfo{year}{2006}).

\bibitem[{\citenamefont{Barford}(2007)\citenamefont{Barford}}]{Barford2007}
\bibinfo{author}{\bibfnamefont{W.}~\bibnamefont{Barford}},
  \bibinfo{journal}{J.\ Chem.\ Phys.} \textbf{\bibinfo{volume}{126}},
  \bibinfo{pages}{134905} (\bibinfo{year}{2007}).

\bibitem[{\citenamefont{Gierschner and Beljonne}(2009)}]{Gierschner2008}
\bibinfo{author}{\bibfnamefont{J.}~\bibnamefont{Gierschner}} 
 \bibnamefont{et~al.},
  \bibinfo{journal}{J.\ Chem.\ Phys.} \textbf{\bibinfo{volume}{130}},
  \bibinfo{pages}{044105}
  (\bibinfo{year}{2009}).
  
  \bibitem[{\citenamefont{Wolffs et~al.}(2008)\citenamefont{Wolffs, Korevaar, Jonkheijm, Henze, Feast, Schenning, and Meijer}}]{Wolffs08}
\bibinfo{author}{\bibfnamefont{M.}~\bibnamefont{Wolffs}}
 \bibnamefont{et~al.},
  \bibinfo{journal}{Chem. Comm.} \textbf{\bibinfo{volume}{38}},
  \bibinfo{pages}{4613} (\bibinfo{year}{2008}).

\bibitem[{\citenamefont{Henze et~al.}(2006)\citenamefont{Henze, Feast,
  Gardebien, Jonkheijm, Lazzaroni, Leclere, Meijer, and Schenning}}]{Henze2006}
\bibinfo{author}{\bibfnamefont{O.}~\bibnamefont{Henze}}
 \bibnamefont{et~al.},
  \bibinfo{journal}{J. Am. Chem. Soc.} \textbf{\bibinfo{volume}{128}},
  \bibinfo{pages}{5923} (\bibinfo{year}{2006}).

\bibitem[{\citenamefont{Spano et~al.}(2007)\citenamefont{Spano, Meskers,
  Hennebicq, and Beljonne}}]{Spano2007a}
\bibinfo{author}{\bibfnamefont{F.~C.} \bibnamefont{Spano}}
 \bibnamefont{et~al.},
  \bibinfo{journal}{J. Am. Chem. Soc.} \textbf{\bibinfo{volume}{129}},
  \bibinfo{pages}{7044} (\bibinfo{year}{2007}); \textbf{\bibinfo{volume}{129}},
  \bibinfo{pages}{16278} (\bibinfo{year}{2007}).

\bibitem[{\citenamefont{Spano et~al.}(2009)\citenamefont{Spano, Clark,
  Silva, and Friend}}]{Spano2009}
\bibinfo{author}{\bibfnamefont{F.~C.} \bibnamefont{Spano}}
 \bibnamefont{et~al.},
  \bibinfo{journal}{J.\ Chem.\ Phys.} \textbf{\bibinfo{volume}{130}},
  \bibinfo{pages}{074904} (\bibinfo{year}{2009}).
  
\bibitem[{\citenamefont{Spano et~al.}(2008)\citenamefont{Spano, Meskers,
  Hennebicq, and Beljonne}}]{Spano2008}
\bibinfo{author}{\bibfnamefont{F.~C.} \bibnamefont{Spano}}
 \bibnamefont{et~al.},
  \bibinfo{journal}{J. Chem. Phys.} \textbf{\bibinfo{volume}{129}},
  \bibinfo{pages}{024704} (\bibinfo{year}{2008})
  
  
  \bibitem{SuppInfo}
See EPAPS document no.\ XXX for fluence-dependent measurements. This document can be reached through a direct link in the online article's HTML reference section or via the EPAPS homepage (\verb@http://www.aip.org/pubservs/epaps.html@).






\bibitem[{\citenamefont{Daniel et~al.}(2005)\citenamefont{Daniel, Makereel,
  Herz, Hoeben, Jonkheijm, Schenning, Meijer, Friend, and Silva}}]{Daniel:2005}
\bibinfo{author}{\bibfnamefont{C.}~\bibnamefont{Daniel}}
 \bibnamefont{et~al.},
  \bibinfo{journal}{J. Chem. Phys.} \textbf{\bibinfo{volume}{123}},
  \bibinfo{pages}{084902} (\bibinfo{year}{2005}).



  
\bibitem[{\citenamefont{Psiachos and Mazumdar}(2009)}]{Psiachos2009}
\bibinfo{author}{\bibfnamefont{D.} \bibnamefont{Psiachos}} \bibnamefont{and}
  \bibinfo{author}{\bibfnamefont{S.} \bibnamefont{Mazumdar}},
  \bibinfo{journal}{Phys.\ Rev.\ B} \textbf{\bibinfo{volume}{79}},
  \bibinfo{pages}{155106} (\bibinfo{year}{2009}).


\bibitem[{\citenamefont{Arkhipov et~al.}(2004)\citenamefont{Arkhipov, Bassler, Emelyanova, Hertel, Gublinas, and Rothberg}}]{Arkhipov2004}
\bibinfo{author}{\bibfnamefont{V.}~\bibnamefont{Arkhipov}},
 \bibnamefont{et~al.},
\bibinfo{journal}{Macromol.\ Symposia} \textbf{\bibinfo{volume}{212}},
\bibinfo{pages}{13--24} (\bibinfo{year}{2004}).



\bibitem[{\citenamefont{Scholes}(2008)}]{Scholes2008}
\bibinfo{author}{\bibfnamefont{G.~D.}~\bibnamefont{Scholes}}, 
\bibinfo{journal}{ACS Nano} \textbf{\bibinfo{volume}{2}}, \bibinfo{pages}{523--537}
  (\bibinfo{year}{2008}).

\bibitem{IndoNote}
J.\ Ridley and M.~C.\ Zerner, Theor.\ Chim.\ Acta \textbf{32}, 111--134 (1974). We used a screened Mataga-Nishmoto potential to describe electron-electron interactions in organic condensed phase, see Y.-S. Huang et al. Nature Materials \textbf{7}, 483--489 (2008) and references therein. Note that: (i) this potential places the lowest CT excited state $\sim0.3$\,eV above the lowest exciton state, which compares favorably with the measured energy separation of $\sim 0.38$\,eV in T6 single crystals (M.~A.\ Loi et al., Phys.\ Rev.\ B \textbf{66}, 113102 (2002)); and (ii) the method is probably less accurate for the description of long-range charge-separated states.


\bibitem[{\citenamefont{Barbara et~al.}(2002)\citenamefont{Barbara, Meyer, and Ratner}}]{Barbara1996}
\bibinfo{author}{\bibfnamefont{P.~F.}~\bibnamefont{Barbara}},
\bibinfo{author}{\bibfnamefont{T.~J.}~\bibnamefont{Meyer}},
\bibinfo{author}{\bibfnamefont{M.~A.}~\bibnamefont{Ratner}},
  \bibinfo{journal}{J. Phys. Chem.} \textbf{\bibinfo{volume}{100}},
  \bibinfo{pages}{13148--13168} (\bibinfo{year}{1996}).

\bibitem[{\citenamefont{Silva et~al.}(2002)\citenamefont{Silva, Russell, Dhoot,
  Herz, Daniel, Greenham, Arias, Setayesh, Mullen, and Friend}}]{Silva2002}
\bibinfo{author}{\bibfnamefont{C.}~\bibnamefont{Silva}}
 \bibnamefont{et~al.},
  \bibinfo{journal}{J. Phys.: Condens. Matter} \textbf{\bibinfo{volume}{14}},
  \bibinfo{pages}{9803} (\bibinfo{year}{2002}).


 
 \bibitem[{\citenamefont{Wang et~al.}(2008)\citenamefont{Wang, Mazumdar, Shukla}}]{Wang2008}
\bibinfo{author}{\bibfnamefont{Z.}~\bibnamefont{Wang}},
  \bibinfo{author}{\bibfnamefont{S.}~\bibnamefont{Mazumdar}},
  \bibnamefont{and}
  \bibinfo{author}{\bibfnamefont{A.}~\bibnamefont{Shukla}},
  \bibinfo{journal}{Phys. Rev. B} \textbf{\bibinfo{volume}{78}},
  \bibinfo{pages}{235109} (\bibinfo{year}{2008}).
  
   \bibitem[{\citenamefont{Muntwiler et~al.}(2008)\citenamefont{Muntwiler, Yang, Tisdale, Zhu}}]{Muntwiler2008}
\bibinfo{author}{\bibfnamefont{M.}~\bibnamefont{Muntwiler}},
  \bibnamefont{et~al.},
  \bibinfo{journal}{Phys. Rev. Lett.} \textbf{\bibinfo{volume}{101}},
  \bibinfo{pages}{196403} (\bibinfo{year}{2008}).


\end{thebibliography}

\end{document}